\documentclass[aps,pre,reprint,nofootinbib]{revtex4-1}

\usepackage{amsmath}
\usepackage{amssymb}
\usepackage{amsfonts}
\usepackage{amsthm}
\usepackage{bm}

\usepackage{graphicx}

\newcommand{\Ref}[1]{Ref.~\cite{#1}}

\newcommand{\SEC}[2]{\section{\label{sec:#1}#2}}
\newcommand{\Sec}[1]{Sec.~\ref{sec:#1}}
\newcommand{\FIG}[2]{\caption{\label{fig:#1}#2}}
\newcommand{\Fig}[1]{Fig.~\ref{fig:#1}}
\newcommand{\EQ}[1]{\label{eq:#1}}
\newcommand{\Eq}[1]{Eq.~(\ref{eq:#1})}
\newcommand{\Tbl}[1]{Table~\ref{tab:#1}}

\DeclareMathOperator{\sign}{sign}
\newcommand{\pnt}[1]{\bm{#1}}
\newcommand{\avg}[1]{\langle#1\rangle}
\newcommand{\prm}[1]{#1^\prime}

\newcommand{\const}{\mathrm{const}}

\newcommand{\Lvl}{\mathrm{i} \mathcal{L}}

\begin{document}

\title{Nonequilibrium Langevin dynamics: a demonstration study of shear flow fluctuations in a simple fluid}
\date{\today}
\author{Roman Belousov}\email{belousov.roman@gmail.com}
\affiliation{The Rockefeller University, New York 10065, USA}
\author{E.G.D. Cohen}\email{egdc@mail.rockefeller.edu}
\affiliation{The Rockefeller University, New York 10065, USA}
\affiliation{Department of Physics and Astronomy, The University of Iowa, Iowa
	City, Iowa 52242, USA}
\author{Lamberto Rondoni}\email{lamberto.rondoni@polito.it}
\affiliation{Dipartimento di Scienze Matematiche and Graphene@Polito Lab Politecnico di Torino - Corso Duca degli Abruzzi 24, 10125, Torino, Italy}
\affiliation{INFN, Sezione di Torino - Via P. Giuria 1, 10125, Torino, Italy}
\affiliation{Kavli Institute for Theoretical Physics China, Chinese Academy of Sciences, Beijing 100190, China}
\affiliation{Malaysia Italy Centre of Excellence for Mathematical Sciences, University Putra Malaysia, 43400 Serdang, Selangor, Malaysia}

\begin{abstract}
	The present study is based on a recent success of the second-order stochastic
	fluctuation theory in describing time autocorrelations of equilibrium and nonequilibrium
	physical systems. In particular, it was shown to yield values of the related deterministic
	parameters of the Langevin equation for a Couette flow in a microscopic Molecular Dynamics
	model of a simple fluid. In this paper we find all the remaining constants of the stochastic
	dynamics, which is then numerically simulated and directly compared with the original
	physical system. By using these data, we study in detail the accuracy and precision
	of a second-order Langevin model for nonequilibrium physical systems, theoretically
	and computationally. In addition, an intriguing relation is found between an applied
	external force and cumulants of the resulting flow fluctuations. This is characterized
	by a linear dependence of {\it athermal cumulant ratio}, a new quantity introduced
	here.
\end{abstract}
\keywords{keywords}
\maketitle

\SEC{intro}{Introduction}

The Langevin dynamics, originally inspired by the problem of Brownian motion \cite[Chapters 1-2]{Coffey2012Langevin},
has now become the fundamental stochastic model of fluctuations for equilibrium physical
systems at mesoscopic scales \cite{Nyquist1928,OM1953,MO1953}. Its generalization
to nonequilibrium steady states, though, is still actively developed, as suggested
by a number of recent publications \cite{KSSH2015I,KSSH2015II,MorgadoQ2016,Queiros2016,PRE2016,PRE2016II,PRE2016III}.
A common objective of these studies is to provide a statistical account of an externally
applied force together with its spontaneous variations, which manifest themselves
in fluctuations of the resulting conjugate current in a system of interest.

In our most recent paper \cite{PRE2016III} we showed, that a second-order Langevin
equation, suggested in Ref.~\cite{MO1953}, provides excellent means for quantitative
studies of fluctuations at mesoscopic scales in both, equilibrium and nonequilibrium,
steady-state systems. In particular, it yields an accurate analytical model of the
time autocorrelation function for currents, which was successfully applied to evaluate
the Green-Kubo formula for a transport coefficient. For a general fluctuating quantity $\alpha(t)$,
the Langevin equation of the second order in time, $t$, reads:
\begin{equation}\EQ{lgv1}
	\ddot\alpha(t) + a \dot\alpha(t) + b^2 \alpha(t) = r(t),
\end{equation}
where $a > 0$, $b > 0$ are constants, while $r(t)$ is a random noise.

In the macroscopic limit, \Eq{lgv1} transforms into a deterministic equation \cite[Sec. 2-3]{OM1953}
and, therefore, $r(t)$ must then in general become a constant, $f \in \mathbb{R}$, so that
we have
\begin{equation}\EQ{mac1}
	\ddot\alpha(t) + a \dot\alpha(t) + b^2 \alpha(t) = f,
\end{equation}
which is a differential equation for a damped harmonic oscillator, subject to an
externally applied macroscopic force $f$. Furthermore, its solution for $\alpha(t)$
converges with time to the steady-state ensemble average of \Eq{lgv1}, e.g. \cite{PRE2016III},
with $\avg{\alpha(t)} = f/b^2$ and $\avg{\dot\alpha(t)} = 0$. By comparing Eqs.~(\ref{eq:lgv1})
and (\ref{eq:mac1}), one can see that, at the mesoscopic scales, $r(t)$ represents
the external force, $f = \avg{r(t)}$, as well as its spontaneous variations \cite{OM1953}.

In the equilibrium regime \cite{Nyquist1928,OM1953,PRE2016II}, $f=0$, the stochastic
term $r(t)$ represents effects of the thermal fluctuations and is commonly given
by $r(t) = A \omega(t)$, where $A > 0$ is a constant proportional to the square root
of the system's temperature, while $\omega(t)$ is a Gaussian white noise with zero
mean and unit variance parameters. In general, this model yields a dynamics, which
is accurate up to the third-order statistics, due to symmetry considerations \cite{PRE2016II,PRE2016III}.

Experiments and molecular dynamics simulations, e.g. Refs.~\cite{PRE2016,PRE2016II},
provide an evidence that, besides the thermal fluctuations, an additional source of spontaneous
variations, $\epsilon$, is present in the nonequilibrium (NE) regime, $f\ne0$,
\begin{equation}\EQ{rNE}r^\mathrm{NE}(t) = A \omega(t) + B \epsilon(t/\tau),\end{equation}
where $B \in \mathbb{R}$ and $\tau > 0$ are constants, discussed later. Within the
white noise approximation, essentially two models of this {\it athermal} contribution
were suggested. One is formulated through a Poisson process \cite{KSSH2015I,KSSH2015II,MorgadoQ2016},
and another uses simple exponential noise \cite{PRE2016II}.

When the athermal fluctuations assume values in a continuous domain, rather than
a discrete one \cite{KSSH2015I,KSSH2015II}, the first of the above models is given
by exponential shot noise; cf. Refs.~\cite{MorgadoQ2016,PRE2016II}. Physically, it
can be interpreted as an external force, applied in the form of discrete impulses.
Their magnitude is distributed exponentially with the scale parameter $B$, while their
number, imparted per unit time, is determined by the Poisson law with the rate parameter
$\tau^{-1}$; cf. \Eq{rNE}. In contrast, the second model, based on simple exponential
noise, describes a nonequilibrium force undergoing continuous variations in time; cf. \Ref{PRE2016II}.

What seems unnoticed so far in the discussions of the Langevin dynamics with athermal
noise, is that a nonequilibrium source of fluctuations in $r^\mathrm{NE}(t)$ does
not exclude a possible presence of an additional deterministic constant term, $F$,
on the right hand side of \Eq{lgv1}, which then reads
\begin{equation}\EQ{lgv2}
	\ddot\alpha(t) + a \dot\alpha(t) + b^2 \alpha(t) = F + A\omega(t) + B\epsilon(t/\tau).
\end{equation}
For both models of athermal noise $\epsilon(t/\tau)$ mentioned above, the macroscopic limit
of the right-hand side of \Eq{lgv2} yields $f = F + B/\tau$; cf. \Eq{mac1} and Refs.~\cite{PRE2016II,PRE2016III}.
The necessity of the constant term $F$ in \Eq{lgv2} is demonstrated in \Sec{mic},
where we study in detail nonequilibrium aspects of the fluctuations of $\alpha(t)$.

The purpose of this paper is to demonstrate a complete Langevin representation of
a nonequilibrium physical system. To do this, in \Sec{mic} we first determine all
the parameters of \Eq{lgv2} for the microscopic model of a shear flow in a simple
fluid, which was already studied in \Ref{PRE2016III}. Then, in \Sec{mes} we perform simulations of
the Langevin dynamics and compare their results with the original system.

In our computational study, the two models of athermal noise, discussed earlier,
are treated separately. They generate qualitatively very similar trajectories of
$\alpha(t)$, as observed in stochastic simulations of \Sec{mes}. Although exponential shot
noise \cite{MorgadoQ2016} has discrete singularities, which make its physical interpretation
distinct from simple exponential noise \cite{PRE2016II}, in the second-order Langevin
dynamics this affects only $\dot\alpha(t)$.

In addition, Appendix~\ref{sec:apx} reveals a mathematical analogy between the models
of athermal noise, considered here. In particular, they can be regarded as Pad\'{e}
approximants for certain generating functions of the exact probability distribution,
associated with $\epsilon(t/\tau)$. The mathematical structure, described in Appendix~\ref{sec:apx},
is applicable also to other families of approximations. This, in principle, allows formulation
of alternative models for athermal noise.

Finally, for simulations of the Langevin dynamics, Appendix~\ref{sec:alg} proposes
an algorithm, which is used to integrate \Eq{lgv2} in \Sec{mes}. Our numerical scheme
minimizes sampling of random variables, which is usually a most intensive part of
the computations. In this regard, the Langevin equation with simple exponential noise
has an advantage over exponential shot noise, because it requires only one random
number generation per step of numerical integration; cf. Appendix~\ref{sec:alg}.
In the macroscopic limit our simulation method coincides with a second-order symplectic
algorithm, which respects the time reversibility of \Eq{mac1}.

\SEC{mic}{Molecular Dynamics Simulations}

\begin{figure}
\includegraphics[width=1\columnwidth]{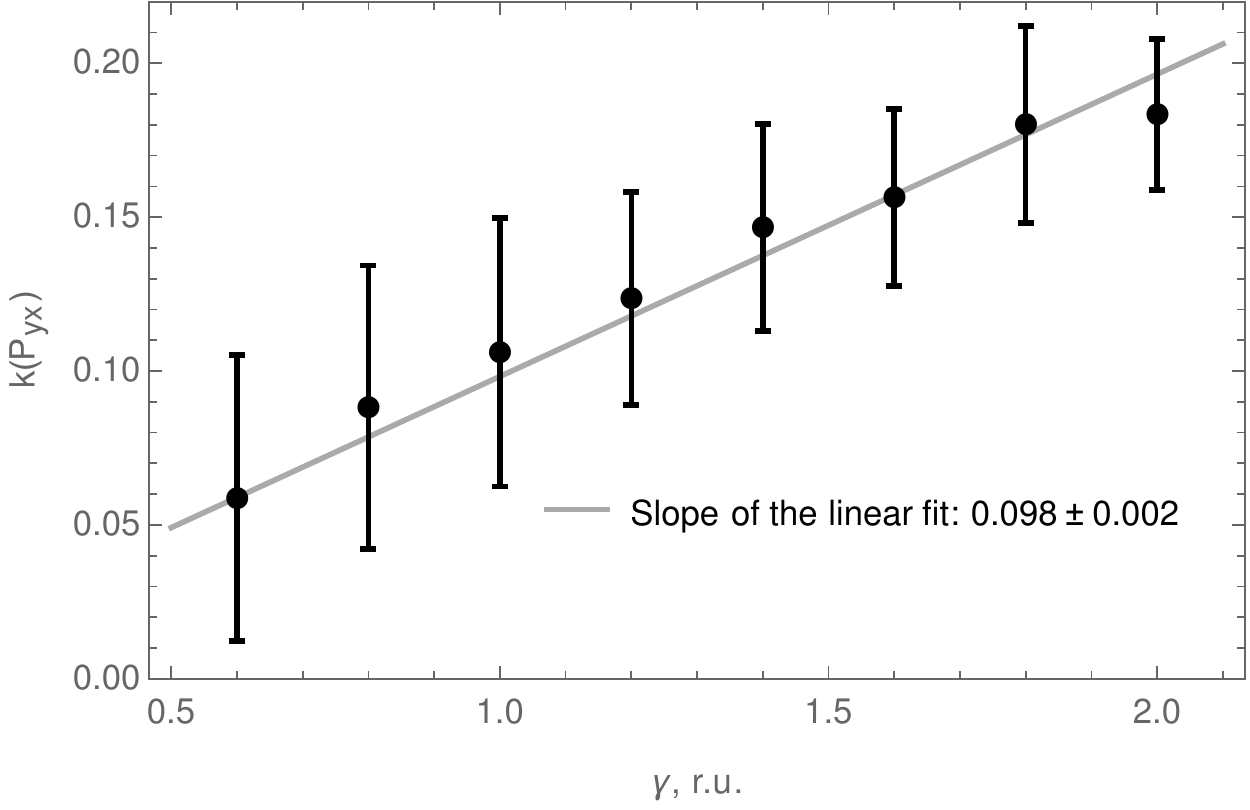}
\FIG{kfig}{
	Plot of the athermal cumulant ratio, $k(P_{yx})$, as a function of the shear rate,
	$\gamma$, for our MD simulations. Error bars are given by one standard deviation.
	A nearly linear trend can be fit to a line, $k(P_{yx}) = \const \gamma$.
}
\end{figure}

\begin{table*}
\caption{\label{tab:stat} Comparison of statistics, in reduced units, between our MD and LD
	simulations for the shear current $\alpha(t)=P_{yx}$. We implemented separately two LD
	models, based on (ex) athermal white exponential noise, and (sh) athermal white exponential
	shot noise.
}
\begin{ruledtabular}
\begin{tabular}{lrrr}
	& LD(ex) & LD(sh) & MD \\
\hline
Mean, $\kappa_1(\alpha)$
& $-1.515 \pm 0.004$ & $-1.518 \pm 0.004$ & $-1.521 \pm 0.004$ \\
Variance, $\kappa_2(\alpha)$
& $ 0.185 \pm 0.003$ & $ 0.185 \pm 0.003$ & $ 0.184 \pm 0.003$ \\
Skewness, $\kappa_3(\alpha)/\kappa_2(\alpha)^{3/2}$
& $ -0.11 \pm  0.03$ & $ -0.16 \pm  0.03$ & $ -0.14 \pm  0.03$ \\
Excess kurtosis, $\kappa_4(\alpha)/\kappa_2(\alpha)^2$
& $  0.15 \pm  0.09$ & $  0.19 \pm  0.07$ & $  0.08 \pm  0.07$ \\
\end{tabular}
\end{ruledtabular}
\end{table*}

In this section we continue to study the same Molecular Dynamics (MD) model of a
shear flow in a simple fluid, as we did in \Ref{PRE2016III}. Here we briefly summarize,
that we consider $N$ particles of equal mass $m$, interacting through the Weeks-Chandler-Anderson
(WCA) potential \cite{WCA} in three dimensions. A constant shear rate $\gamma$, which
is the nonequilibrium force driving a current of linear momentum, is applied by the
SLLOD equations of motion with Lees-Edwards boundary conditions \cite[Chapter 6]{EvansMorriss}.
A constant temperature was maintained by the Nos\'{e}-Hoover (NH) thermostat \cite[Chapter 6]{FrenkelSmit}.

The current in our MD system is given by the $yx$-pressure tensor component, $P_{yx}$,
which expresses a flow of $x$ momentum in the $y$-direction \cite[Sec. 3.8]{EvansMorriss}:
$$ P_{yx} = V^{-1} \sum_{i=1}^{N} (p_{yi} p_{xi} /m + y_i F_{xi}), $$
where $V$ is the volume of the MD simulation cell, while, for the $i$-th particle, $p_{yi}$,
$p_{xi}$, $y_i$, and $F_{xi}$ are, respectively, the $x$ and $y$ components of {\it peculiar}
linear momentum, $y$-coordinate, and $x$-component of the force due to interactions with the
other particles.

Data and results of our simulations are reported in the reduced units (r.u.), described
in \cite[Appendix C]{PRE2016III}. Here we are interested in a nonequilibrium Langevin
equation~(\ref{eq:lgv2}) for $\alpha(t) = P_{yx}$ accurate up to the third-order
moment, which is most reliably estimated for the steady-state probability distribution
of $P_{yx}$ observed in systems of a small size; cf. \Ref{PRE2016}. Therefore, we
conduct our simulations for $N=125$ particles, although at the same number density
$0.8$ r.u. and temperature $1$ r.u., as in \Ref{PRE2016III}.

As shown in \Ref{PRE2016III}, the constants $a$ and $b$ in \Eq{lgv2} do not depend
on $\gamma$ and can be estimated from measurements of the time autocorrelation function
of $P_{yx}$. Also, since the temperature is fixed in our simulations by the NH thermostat,
the parameter $A$ can be determined from the variance for $P_{yx}$, measured in the
equilibrium simulations, i.e. $\kappa_2(P_{yx}|\gamma=0)$; cf. \Sec{intro}.
Hence in this section we are mainly concerned with the remaining parameters of \Eq{lgv2},
$F$, $B$ and $\tau$, which can be obtained, by fitting the first three cumulants
of $P_{yx}$, $\kappa_i(P_{yx}), i=1,2,3$, in a nonequilibrium steady-state $\gamma\ne0$,
as follows.

Formulas for the steady-state cumulants of $\alpha(t)$, which evolves according to
\Eq{lgv2}, can be derived, by using the method of Ref.~\cite[Appendix A]{PRE2016III}.
For this we need to consider a random variable $R(t)$, given by the time integral
of the force terms in the Langevin dynamics:
\begin{equation}\EQ{Rt} R(t) = \int_0^{t} d\prm{t} [F + A \omega(\prm{t}) + B \epsilon(\prm{t})]. \end{equation}
The steady-state cumulants of $\alpha(t)$ can be then expressed as
\begin{eqnarray}\EQ{kappar}
	\kappa_1(\alpha) &=& \frac{\kappa_1(R)}{b^2 t};\quad
	\kappa_2(\alpha) = \frac{\kappa_2(R)}{2 a b^2 t};\nonumber\\
	\kappa_3(\alpha) &=& \frac{2 \kappa_3(R)}{3 b^2 (2 a^2 + b^2) t}.
\end{eqnarray}
cf. Ref.~\cite[Appendix A]{PRE2016III}.

Equation~(\ref{eq:kappar}) can be further expanded, once the form of athermal noise in \Eq{lgv2},
$B \epsilon(t/\tau)$, is specified. We discussed in \Sec{intro}, that in this paper we consider
separately two models, simple exponential noise ($\mathrm{ex}$) and exponential shot noise ($\mathrm{sh}$).
From now on, to distinguish them we will use subscripts in the parameters of \Eq{lgv2},
respectively, $F_\mathrm{ex} + B_\mathrm{ex} \epsilon_\mathrm{ex}(t/\tau_\mathrm{ex})$
and $F_\mathrm{sh} + B_\mathrm{sh} \epsilon_\mathrm{sh}(t/\tau_\mathrm{sh})$. In
the case of simple exponential noise, \Eq{kappar} thus yields:
\begin{eqnarray}\EQ{kappaex}
	\kappa_1(\alpha) &=& \frac{F_\mathrm{ex} + B_\mathrm{ex}/\tau_\mathrm{ex}}{b^2};\quad
	\kappa_2(\alpha) = \frac{A^2 + B_\mathrm{ex}^2/\tau_\mathrm{ex}}{2 a b^2};\nonumber\\
	\kappa_3(\alpha) &=& \frac{4 B_\mathrm{ex}^3}{3 b^2 (2 a^2 + b^2) \tau_\mathrm{ex}}.
\end{eqnarray}
And similarly, for exponential shot noise, we get
\begin{eqnarray}\EQ{kappash}
	\kappa_1(\alpha) &=& \frac{F_\mathrm{sh} + B_\mathrm{sh}/\tau_\mathrm{sh}}{b^2};\quad
	\kappa_2(\alpha) = \frac{A^2 + 2 B_\mathrm{sh}^2/\tau_\mathrm{sh}}{2 a b^2};\nonumber\\
	\kappa_3(\alpha) &=& \frac{4 B_\mathrm{sh}^3}{b^2 (2 a^2 + b^2) \tau_\mathrm{sh}}.
\end{eqnarray}
cf. \cite{PRE2016II,PRE2016III}. The constants $A$, $a$ and $b$ do not depend on the
shear rate, and we also have
\begin{equation}\EQ{eqkappa2} \kappa_2(\alpha|\gamma=0) = \frac{A^2}{2 a b^2}; \end{equation}
cf. \Sec{intro}. Equations~(\ref{eq:kappaex})-(\ref{eq:eqkappa2}) suggest then
to define a nonequilibrium part of the second cumulant as
$$ \chi(\alpha) = \kappa_2(\alpha) - \frac{A^2}{2 a b^2}. $$

\begin{figure*}
\includegraphics[width=2\columnwidth]{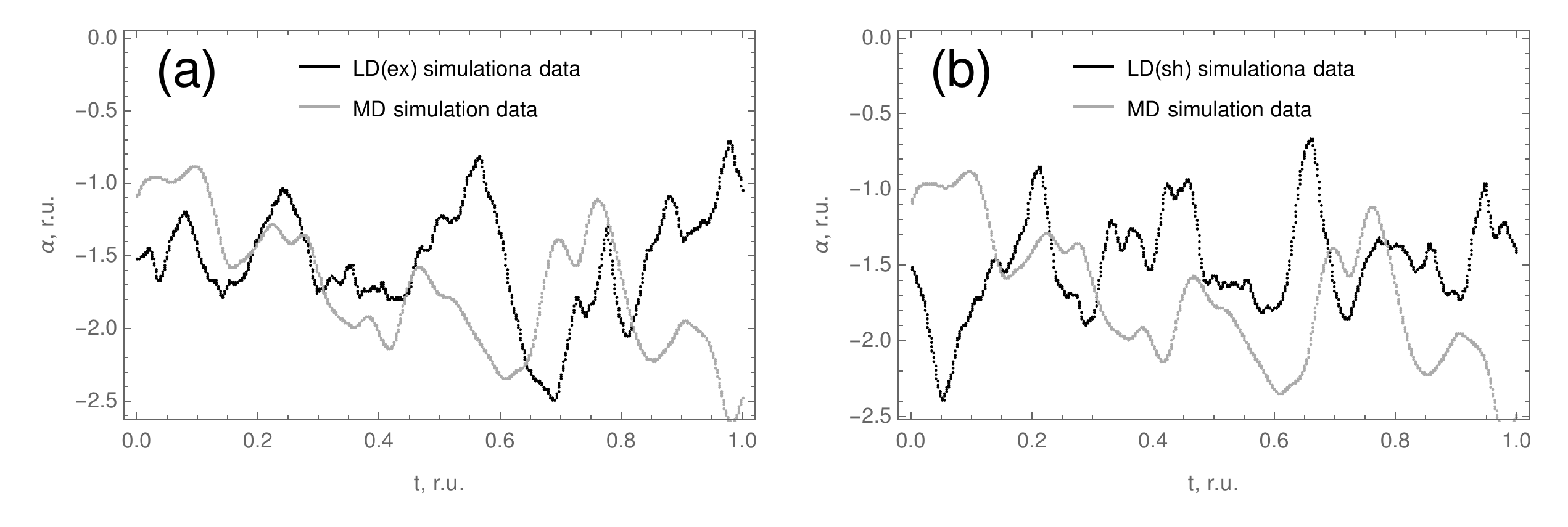}
\FIG{trj}{
	Qualitative comparison of sample trajectories for $\alpha(t)=P_{yx}$ in our MD
	and LD simulations: (a) the MD simulation vs the LD model with simple exponential noise (ex);
	(b) the MD simulation vs the LD model with exponential shot noise (sh).
}
\end{figure*}

Formulas~(\ref{eq:kappaex}) and (\ref{eq:kappash}) form a system of equations, which
can be solved for the parameters of \Eq{lgv2},
\begin{eqnarray}\EQ{kappas}
	B_\mathrm{ex} &=& \frac{3 (2 a^2 + b^2) \kappa_3(\alpha)}{8 a \chi(\alpha)} = 3 B_\mathrm{sh} / 2;
	\nonumber\\
	\tau_\mathrm{ex} &=& \frac{9 (2 a^2 + b^2)^2 \kappa_3^2(\alpha)}{128 a^3 b^2 \chi^3(\alpha)} = 9 \tau_\mathrm{sh} / 8;
\end{eqnarray}
whereas $F$ is calculated by the residual principle, $F = b^2\kappa_1(\alpha) - B/\tau$, with
$B$ and $\tau$ being the parameters of the chosen athermal noise model.

In \Sec{intro} we promised to show, that the parameter $F$ is essential in \Eq{lgv2}.
In particular, it allows our Langevin model for $P_{yx}$ to achieve the statistical
accuracy of third order. Consider the following quantity, which relates the first
three cumulants of $P_{yx}$,
$$ k(P_{yx}) = \frac{\chi^2(P_{yx})}{\kappa_1(P_{yx}) \kappa_3(P_{yx})}, $$
Below we refer to $k(P_{yx})$ as {\it the athermal cumulant ratio}, because it captures
effects of the athermal noise in \Eq{lgv2} and vanishes in equilibrium systems together
with $\chi(P_{yx})$; cf. \Eq{eqkappa2}.

As can be deduced from Eqs.~(\ref{eq:kappaex}) and (\ref{eq:kappash}), the athermal
cumulant ratio, for both models of athermal noise considered in \Eq{lgv2}, has a
general form
\begin{equation}\EQ{kform}
	k \propto (\const_1 F \tau / B + \const_2)^{-1}{,}
\end{equation}
where $\const_1$ and $\const_2$ are some constants, which do not depend on the shear
rate. By substituting $F=0$ into \Eq{kform}, we see, that without a force term $F$
on the right-hand side of \Eq{lgv2} in both cases, simple exponential noise and exponential shot noise,
$k$ would be independent of $\gamma$.

We plot $k(P_{yx})$ as a function of $\gamma$ for our MD simulations in \Fig{kfig},
where a nearly linear trend can be observed. In particular, this implies, that $\chi(P_{yx})$,
which essentially expresses the athermal variance of $P_{yx}$, vanishes faster than
the mean and asymmetry of its steady-state probability distribution, respectively,
$\kappa_1(P_{yx})$ and $\kappa_3(P_{yx})$.  Such behavior of athermal statistics
is consistent with the observation, that the fluctuations of the nonequilibrium force
become negligible in the near-equilibrium systems \cite{PRE2016}, i.e. $\chi(P_{yx})\approx0$
and $\kappa_3(P_{yx})\approx0$, and \Eq{lgv2} thus can be approximated by
$$\ddot\alpha(t) + a \dot\alpha(t) + b^2 \alpha(t) \approx f + A\omega(t),$$
which agrees in the macroscopic limit with \Eq{mac1}, cf. \Sec{intro}.

In \Eq{kform}, the linear dependence of the athermal cumulant ratio on the shear
rate can only arise due to the term proportional to $F$, which therefore can not vanish.
In fact, \Eq{lgv2} admits {\it an alternative form}, in which one of its parameters
is related linearly to the athermal cumulant ratio, as shown in Appendix~\ref{sec:acr}.

As anticipated in \Sec{intro}, \Eq{lgv2}, with $\epsilon$ represented by white exponential
noise or white exponential shot noise, is accurate up to the third-order statistics.
For this, three parameters, $F$, $B$, and $\tau$, are required to fit exactly three
cumulants $\kappa_1$, $\kappa_2$ and $\kappa_3$; cf. Eqs.~(\ref{eq:kappaex})-(\ref{eq:kappas}).

\SEC{mes}{Langevin Dynamics Simulation}

Once all the parameters of the stochastic differential equation~(\ref{eq:lgv2}) are found,
it can be simulated numerically. In Appendix~\ref{sec:alg} we design an integration algorithm,
which is consistent with the time reversibility of the Langevin Dynamics (LD) in the macroscopic
limit, \Eq{mac1}. Below we report the results of simulations, obtained with this numerical
scheme.

As described in \Sec{mic}, we determined the parameters of \Eq{lgv2} for the shear current
$P_{yx}$ and the two models of athermal noise, $\epsilon_\mathrm{ex}(t)$ and $\epsilon_\mathrm{sh}(t)$,
from our MD calculations, obtained at $\gamma = 1$ r.u. \Tbl{stat} summarizes results of our
simulations. Statistics of the original MD data are reproduced very accurately and
precisely by our numerical LD model up to the third-order statistics.

A notable discrepancy between the simulations is observed in \Tbl{stat} only for the excess
kurtosis. This should be expected, because the parameters of \Eq{lgv2} were obtained
by fitting solely the first three cumulants of the MD data, and the two LD models,
considered here, differ by their fourth cumulants in principle. Such level of precision,
however, is beyond the statistical and numerical resolution of our simulations. Indeed,
the magnitude of the excess kurtosis for the MD data is comparable to its standard
deviation. Its theoretical value in our LD models is $17\cdot10^{-5}$ for the case
of simple exponential noise, and $12\cdot10^{-5}$ for the case of exponential shot
noise. These predictions of the excess kurtosis are also much less than the respective
standard deviations in \Tbl{stat}. Therefore, within the statistical uncertainties,
the excess kurtosis of our MD and LD models can effectively be considered zero,
i.e. Gaussian-like.

To integrate numerically \Eq{lgv2}, we used the same time step, $\Delta{t} = 10^{-3}$,
as in the original MD simulations; cf. \Ref{PRE2016III}. Therefore, we can compare
sample trajectories, which are traced pointwise in time for our LD models and the
original MD system in \Fig{trj}. Both, simple exponential noise and exponential shot
noise, produce a qualitatively similar behavior of $\alpha(t)$. In particular, the
two stochastic models match closely the amplitude of $\alpha(t)$ fluctuations, as
well as the time scale of their onset and decay, observed in the MD simulations.

The trajectory of the original MD system in \Fig{trj} mostly resembles those of the
LD simulations, perhaps, except for one subtle detail. Namely, on short time intervals,
the stochastic dynamics of the time derivative $\dot\alpha(t)$ may undergo quick alterations,
which makes the trajectory of $\alpha(t)$ look noisy. This is especially conspicuous
for $\dot\alpha(t)\approx0$, where this behavior leads to quick alterations of sign
$\dot\alpha(t)\gtrless0$ with a nearly constant value of $\alpha(t)$, like in \Fig{trj}(a)
around $0.35<t<0.45$ r.u. or in \Fig{trj}(b) $0.50 < t < 0.55$. In contrast, the
trajectory of our MD simulation appears more resistant to changes of its direction, $\dot\alpha(t)$,
and thus slightly smoother.

The alterations of $\dot\alpha(t)$, mentioned above, might be an artifact of the
white noise approximation. In principle, a colored noise, which introduces additional
correlations into the stochastic forces on the right-hand side of \Eq{lgv2}, e.g.
\cite{Queiros2016}, could make the dynamics of $\dot\alpha(t)$ more inertial. Possibly
this would make the trajectory of $\alpha(t)$ smoother, by reducing the chance of
quick sign alterations in its time derivative.

\SEC{final}{Conclusion}

In this paper we have demonstrated, how efficiently the Langevin dynamics can represent
certain nonequilibrium systems in a steady state. Section~\ref{sec:mic} describes
a way to find the parameters of such representation from the statistics of the original
system. The numerical simulations of \Eq{lgv2}, thus obtained, and the integration
algorithm, proposed in Appendix~\ref{sec:alg}, yield results, which are both, statistically
and qualitatively, very accurate and precise; cf. \Sec{mes}.

Two models of athermal noise for \Eq{lgv2}, i.e. simple white exponential noise and white
exponential shot noise, were compared in \Sec{mes}. Within the third-order statistics they
provide equivalent results. As explained in Appendix~\ref{sec:apx}, these two models belong to two,
among many other, possible families of approximants for certain generating functions of the
true probability distribution, associated with athermal noise $\epsilon(t)$. Therefore, from
the statistical and mathematical perspectives, they are equivalent and interchangeable. The
difference of the physical interpretations between the considered models of athermal noise,
which might have distinct consequences for the first-order Langevin dynamics \cite{MO1953,PRE2016II},
in the second-order equation~(\ref{eq:lgv2}) applies mostly to $\dot\alpha(t)$, and does not
affect significantly the behavior of $\alpha(t)$.

The above models of athermal noise differ by their values of the fourth cumulant. Within the
statistical uncertainties of the original system considered in \Sec{mic}, this discrepancy
is, however, completely negligible; cf \Sec{mes}. Statistically the excess kurtosis of our
MD and LD data for $\alpha(t)$ is comparable to that of a Gaussian probability distribution,
which equals zero.

To characterize the third-order statistics of athermal noise, we introduced a new
quantity, {\it athermal cumulant ratio}, which has a nearly linear dependence on
the external shear rate. This observation was analyzed in \Sec{mic}, to justify the
constant term $F$ in \Eq{lgv2}. Hence, the second-order Langevin representation of
our original model needs, in total, specification of six constants, $a$, $b$, $A$,
$B$, $\tau$, and $F$. Furthermore, \Eq{lgv2} can be reparametrized so, that the athermal
cumulant ratio is linearly related to one of the new parameters; see Appendix~\ref{sec:acr}.

Finally, at short-time scales we observed in \Sec{mes} some features of the LD simulations,
which appear to be artifacts of the white noise approximation. As discussed there,
a correlated noise could, in principle, improve this aspect of the Langevin equation.
A description of noise correlations may require to add another parameter to the model. This,
however, seems far from being practical, since there would be more unknowns than statistically
significant measurements to fit; cf. Secs.~\ref{sec:mic}-\ref{sec:mes}. Instead, one could try
to construct a first-order Langevin equation with a correlated noise, which would reproduce
the time autocorrelation function for $\alpha(t)$ in \Eq{lgv2}. Indeed, the term $a\dot\alpha(t)$
of the second-order dynamics is required for the accurate description of time autocorrelations
in the modeled system; cf. \Ref{PRE2016III}. This introduces essentially a nuisance parameter $a$,
if we are interested merely in the behavior of $\alpha(t)$. One may hope, however, that the
time correlation function of a first-order Langevin dynamics, with a properly designed colored
noise, would match that of $\alpha(t)$ in \Eq{lgv2}.

\appendix
\SEC{apx}{Models of athermal noise}
In \Ref{PRE2016II} we have presented a derivation of exponential white noise for
the Langevin dynamics of nonequilibrium physical systems, as an alternative to
white exponential shot noise \cite{MorgadoQ2016}. As mentioned in \Sec{intro}, these two models
essentially account for the spontaneous variations of a nonequilibrium force, which are assumed
independent from the thermal fluctuations and, therefore, called {\it athermal} \cite{MorgadoQ2016}.

The derivation of exponential white noise \cite{PRE2016II} followed a procedure similar
to that of the Gaussian white noise in Ref.~\cite[Sec. I.1]{Chandrasekhar1943}, used to model
thermal fluctuations in both, equilibrium and nonequilibrium, systems. This procedure suggests
to design a discrete random walk, which incorporates physically relevant assumptions and properties
of the fluctuating force. Then, a definition of stochastic noise, which can be used in a Langevin
equation, naturally arises in a properly chosen continuous limit.

In this section we revisit the above procedure and show, that it relies on a more
general mathematical structure of asymptotic approximations, which allows to derive
various stochastic noises with desired statistical properties. In fact both, simple exponential
noise and exponential shot noise, belong to two families of random processes, which can be
obtained in this manner.

Let us begin with a review of the Gaussian white noise $\omega(t)$, which is defined
as a time derivative of the Gaussian random process $\Omega(t)$ with the zero mean
and the variance $A^2 t$, so that
$$\Omega(t) = \int_0^{t} d\prm{t} A \omega(\prm{t})\text{,}$$
where $A>0$ is a constant.

In the derivation of Gaussian white noise we consider a random walk, where the walker
moves in a discrete series of steps from an initial position $x_0 = 0$. The length
of each step $\xi\approx A \omega(t) dt$ is an independent random variable with a given probability
density $p(\xi)$. A physically relevant property, which we want to incorporate in this model,
is a spatial symmetry for the forward ($\xi>0$) and backward ($\xi<0$) displacements. Statistically
this means $p(\xi) = p(-\xi)$. The walker's position $x$ after $n$ steps is then given by a
sum of $n$ independent realizations of $\xi$. Therefore, the moment-generating function of
the random variable $x$ equals the moment-generating function of $\xi$ raised to the power
$n$, and thus
\begin{eqnarray}
	\EQ{mu} \mu(\tilde{x}) &=& [\mu(\tilde\xi)]^n\vline_{\tilde\xi=\tilde{x}};\\
	\EQ{kappa} \kappa(\tilde{x}) &=& \ln \mu(\tilde{x}) = n \kappa(\tilde\xi)\vline_{\tilde\xi=\tilde{x}}\text{,}
\end{eqnarray}
where $\tilde\xi$ and $\mu(\tilde\xi)$ are, respectively, the reciprocal dual variable\footnote{
	The moment-generating function, which is the Laplace transform of a probability density,
	i.e. $\avg{\exp(\xi\tilde\xi)}$, depends on a variable $\tilde\xi$, a reciprocal dual of $\xi$.
} of $\xi$ and its moment-generating function, and likewise for $x$, while $\kappa(\cdot)$
stands for the cumulant-generating function. From now on we consistently denote duals of random
variables by the tilde.

By expanding the cumulant-generating function of $\xi$ in Taylor series at $\tilde\xi=0$,
\begin{equation}\EQ{Taylor}
	\kappa(\tilde\xi) = \sum_{i=0}^{n} \kappa_i(\xi) \frac{\tilde\xi^i}{i!}
		= \kappa_2(\xi) \tilde\xi^2 / 2 + \mathcal{O}(\tilde\xi^{4}),
\end{equation}
we use the fact that, for a symmetric random variable, all cumulants of odd orders $\kappa_1(\xi)$,
$\kappa_3(\xi)$, and so on, must vanish, whereas the normalization of a probability density
function always requires $\kappa_0(\cdot)=0$. The Taylor series \Eq{Taylor}, truncated at the
second term is a cumulant-generating function of a Gaussian probability distribution
$$\kappa_G(\tilde\xi) = \kappa_2(\xi) \tilde\xi^2/2,$$
which can be used to approximate $\kappa(\tilde\xi) \approx \kappa_G(\tilde\xi)$ up to the third
order statistics; cf. \Eq{Taylor}. Hence, by virtue of \Eq{kappa} we further obtain,
\begin{equation}\EQ{gaussn}\kappa(\tilde{x}) \approx n \kappa_2(\xi) \tilde{x}^2/2.\end{equation}
Not all orders of truncated power series are valid cumulant-generating functions. The second
order approximation, however, always corresponds to a Gaussian probability distribution.

Notice that we have not invoked the Central Limit Theorem so far. Now we do this,
in order to ensure that \Eq{gaussn} holds asymptotically in the limit of large $n$,
at least for fluctuations of size $\sqrt{n}$. At last, the continuous limit of the
described random walk is taken, by requiring that, in an infinitesimal time $dt$,
the walker makes a number of steps $dn$ with a rate $\tau = dt/dn$, so that \Eq{gaussn}
becomes
\begin{equation}\EQ{gausst}\kappa(\tilde{x}) \approx \frac{\kappa_2(\xi)  t \tilde{x}^2}{2\tau},\end{equation}
which is the cumulant-generating function for the process $\Omega(t)$ with $A^2 = \kappa_2(\xi)/\tau$.

Instead of simply reviewing exponential noise, however, we now derive a more general
family of random processes, to which it pertains. To do this, we again consider a random walk
problem $\xi\approx B\epsilon(t/\tau)dt$, similar to the one described above. Due to the physical
assumptions discussed in \Ref{PRE2016II}, this time we require $\kappa_3(\xi) > 0$ and that
the probability density $p(\xi)$ vanishes for $\xi<0$, which implies $\kappa_1(\xi)\ge0$.

Earlier we used power series to approximate the cumulant-generating function for the symmetric
random walk. This time we will use a Pad\'{e} approximant \cite{BakerEssentials} for the moment-generating
function $\mu(\tilde\xi)$.

Just like with power series, not all orders of the numerator and denominator, $i$
and $j$, respectively, in a Pad\'{e} approximant $\mu_{[i/j]}(\tilde\xi)\approx\mu(\tilde\xi)$,
are valid moment-generating functions. We set $i = j = 1$, which is accurate at least
up to the second-order statistics, for $i+j=2$, and has a general form \cite[Chapter 1]{BakerEssentials}
\begin{equation}\EQ{MuPade1}
	\mu_{[1/1]}(\tilde\xi) = \frac{a_{1,1} + a_{1,2} \tilde\xi}{a_{2,1} + a_{2,2} \tilde\xi},
\end{equation}
where $a_{1,1}$, $a_{1,2}$, etc. are constants. The normalization of the probability
distribution, though, requires that $\mu_{[1/1]}(0)=1$ and, thus, $a_{2,1} = a_{1,1}$.
With $b_1 = a_{1,2}/a_{1,1}$ and $b_2 = a_{2,1}/a_{1,1}$, \Eq{MuPade1} becomes
\begin{equation}\EQ{MuPade2}
	\mu_{[1/1]}(\tilde\xi) = \frac{1 + b_1 \tilde\xi}{1 + b_2 \tilde\xi}
		= \frac{b_1}{b_2} + \frac{1-b_1/b_2}{1+b_2\tilde\xi}
		= q + \frac{1-q}{1-b\tilde\xi}.
\end{equation}
where in the last equality we used another substitution $q=b_1/b_2$ and $B = -b_2$.
The probability density, which corresponds to a moment-generating function $\mu_{[1/1]}(\tilde\xi)$,
is
\begin{equation}
	p_{[1/1]}(\xi) = q \delta(\xi) + (1-q) H[\sign(B) \xi] \exp(-\xi/B)/B,
\end{equation}
where $\delta(\cdot)$ and $H(\cdot)$ are, respectively, the Dirac delta and Heaviside step
functions. The above equation is a valid probability density function $p_{[1/1]}(\xi)>0$ only
for $0 \le q \le 1$. Furthermore, our assumptions $\kappa_1(\xi)>0$ and $\kappa_3(\xi)>0$ impose
another restriction $B>0$. The continuous limit of the described approximation can be obtained
by requiring, that the walker makes steps with a rate $\tau=dt/dn$, just like we did earlier
in the derivation of $\omega(t)$. The result is a defintion of random noise $\epsilon_q(t/\tau)$,
which upon the time integration generates the following stochastic process
$$E_q(t/\tau) = \int_0^t d\prm{t} B \epsilon_q(t/\tau),$$
with the cumulant-generating function and the probability density, respectively,
\begin{eqnarray}\EQ{exp}
	\kappa(\tilde{E}_q) &=& \ln[q + (1-q)/(1-B \tilde{E}_q)] t/\tau,\nonumber\\
	p(\tilde{E}_q) &=& q^{t/\tau} \delta(E_q/B)\nonumber\\
		&+& \sum_{i=0}^{\infty} \binom{t/\tau}{i} q^{t/\tau-i}(1-q)^{i} p_\Gamma(E_q;i,B)\text{,}
		\nonumber\\
\end{eqnarray}
where $p_\Gamma(\cdot;i,B)$ is a probability density of the Gamma distribution with
the shape parameter $i$ and the scale $B$.

The stochastic noise $\epsilon_q(t/\tau)$ reduces to simple exponential noise when $q = 0$,
as it does under the conditions of random walk problem assumed in \Ref{PRE2016II}. The general
case of $q\ne0$ introduces one more parameter into \Eq{lgv2}, which is unpractical within
the statistical precision of our model; cf. Secs. \ref{sec:mic} and \ref{sec:mes}.

A shot noise approximation of the athermal noise \Eq{exp} can be derived in two ways. First, we
may alter slightly the interpretation of a continuous limit for the random walk.
Above the walker attempts $dn = dt / \tau$ steps in an infinitesimal time $dt$ {\it deterministically}.
Instead, we could require, that the walker attempts a step with a probability proportional
to a time interval $dt$, so small that it can make at most one displacement forward with the
probability $dt/\prm{\tau}$, or to remain motionless with a probability $1-dt/\prm{\tau}$.
Given that the length of each step is the random variate $\xi$, this sheme generates a compound
Poisson process, $P_q(t)$; cf. \cite[Sec. 2.3]{KardarI}. Its cumulant-generating function is
\begin{equation}
	\kappa(\tilde{P}_q) = \frac{[\mu(\tilde\xi)-1] t}{\prm{\tau}}\vline_{\tilde\xi=\tilde{P}_q},
\end{equation}
which together with the approximant given by \Eq{MuPade2} yields
\begin{equation}
	\kappa(\tilde{P}_q) \approx \frac{(1-q) B \tilde{P}_q t}{\tau(1-B \tilde{P}_q)}.
\end{equation}
The dimensionless parameter $(1-q)$ can be further absorbed into the constant $\tau = \prm{\tau}/(1-q)$,
and we thus obtain the compound Poisson process $P(t)$ with exponentially distributed intensity
\begin{equation}\EQ{pois}
	\kappa(\tilde{P}) = \frac{B \tilde{P} t}{(1-B x) \tau},
\end{equation}
cf. \Ref{PRE2016II}.

Another way to derive \Eq{pois} is to seek directly a Pad\'{e} approximant for the cumulant
generating function $\kappa(\tilde\xi) \approx \kappa_{[1/1]}(\tilde\xi)$.
Like for $\mu_{[1/1]}$ in \Eq{MuPade1}, we begin with a most general form
\begin{equation}\EQ{KaPade1}
	\kappa_{[1/1]}(\tilde\xi) = \frac{a_{1,1} + a_{1,2} \tilde\xi}{a_{2,1} + a_{2,2} \tilde\xi},
\end{equation}
where $a_{1,1}$, $a_{1,2}$, etc. are constants. Since a cumulant-generating function
must vanish at $\tilde\xi=0$, as required by the normalization of probability densities,
we then have $a_{1,1} = 0$. With $B = -a_{2,2}/a_{2,1}$ and $a_2=-a_{1,2} / (B a_{2,2})$,
\Eq{KaPade1} becomes
\begin{equation}\EQ{KaPade2}
	\kappa_{[1/1]}(\tilde\xi) = \frac{a_2 B \tilde\xi}{1 - B \tilde\xi}.
\end{equation}

Now we can take a continuous limit in the manner we did for the Gaussian and simple exponential
noise, to obtain a process $\prm{P}(t)$
\begin{equation}
	\kappa(\prm{\tilde{P}}) = \frac{a_2 B \prm{\tilde{P}}t}{(1-B \tilde{P}) \prm{\tau}},
\end{equation}
where again the dimensionless constant $a_2$ can be absorbed into $\tau = \prm{\tau}/a_2$
to yield \Eq{pois}.

Finally, we would like to remark, that the Central Limit Theorem ensures that the above approximations
hold asymptotically. Indeed, all the random walk problems considered here tend to
a Gaussian probability distribution for $n\to\infty$. Our Pad\'{e} approximants have
the same asymptotic limit, since they are accurate up to the second-order statistics.

Above we demonstrated, how the Pad\'{e} approximation can be applied to formulate
a random noise with presupposed statistical properties. In principle, however, other
forms of approximants could be used in exactly the same way. For example, \Ref{DanchenkoChunaev2011}
suggests a general approach, which allows to generate various mathematical series,
which fit the Taylor expansion of the approximated function. In particular, one of these
series can be generated from the simple fraction $(1-x)^{-1}$ \cite{DanchenkoChunaev2011}.
Under certain conditions, this yields a hyperexponential approximant for the moment-generating
function $\mu(\tilde\xi)$ \cite{Hyperexp}. In other words, the Central Limit Theorem,
as used in this section, and the approach of \Ref{DanchenkoChunaev2011} give rise
to a multitude of stochastic processes, which can fit various dynamical models.

\SEC{acr}{Athermal cumulant ratio revisited}

In this section of Appendix we would like to show an alternative parameterization of
\Eq{lgv2}, which provides a particularly simple expression for the athermal cumulant
ratio introduced in \Sec{mic}. We again consider the random walk problem, formulated
in Appendix~\ref{sec:apx} for athermal noise. This time, as an approximant of the moment-generating
function, we would like to use the following expression
\begin{equation}\EQ{muqf}\mu_{q,\phi}(\tilde\xi) = \exp[(1-q) \phi \tilde\xi]/(1-q \phi \tilde\xi), \end{equation}
which is a moment-generating function of the exponentially distributed random variate,
shifted by a location parameter $(1-q) \phi$, and corresponds to the probability density
$$ p_{q,\phi}(\xi) = \frac{H[\xi - (1-q) \phi]}{q\phi} \exp\left(-\frac{\xi}{q\phi}\right), $$
with $f\ge0$ and $0\le q \le 1$. The Taylor expansion of the cumulant-generating
function for the this distribution is
$$\kappa_{q,\phi}(\tilde\xi) = \phi\tilde\xi + q^2 \phi^2 \tilde\xi^2/2 + q^3 \phi^3 \tilde\xi^3/6 + O(\xi^4),$$
in which we always can choose the two parameters $\phi=\kappa_1(\xi)$ and $q=\sqrt{\kappa_2(\xi)/\phi^2}$,
so that
\begin{equation}\EQ{cumf} \kappa(\tilde\xi) = \kappa_{q,\phi}(\tilde\xi) + O(\tilde\xi^3).\end{equation}
This equation is inspired by the approach of \Ref{DanchenkoChunaev2011}, which is based on matching
the Taylor expansion coefficients of the approximated function and the approximant; cf. Appendix~{apx}.

In the continuous limit, with the number of steps per unit time $\tau^{-1} = dn/dt$,
\Eq{cumf} yields a cumulant-generating function of a process $E_{q,\phi}(t)$,
$$
	\kappa(\tilde{E}_{q,\phi}) = \frac{t (1-q) \phi \tilde{E}_{q,\phi}}{\tau} - \frac{t\ln(1-q \phi \tilde{E}_{q,\phi})}{\tau}.
$$
If we put $F = (1-q) \phi/\tau$ and $B = q \phi$, the process $E_{q,\phi}(t)$ above can be expressed
as
\begin{equation}\EQ{Efq}
	E_{q,\phi}(t) = \int_0^{t} \prm{dt}[F + \epsilon(\prm{t}/\tau)] \text{,}
\end{equation}
where $\epsilon(t/\tau)$ is simple exponential noise; cf. Sec.~\ref{sec:mic}.
By using \Eq{Efq}, \Eq{lgv2} can be written as
$$\ddot\alpha(t) + a \dot\alpha(t) + b^2 \alpha(t) = A\omega(t) + dE_{q,\phi}(t)/dt.$$

In this new parameterization of the Langevin equation, which replaces constants $B$ and $F$
by $q$ and $\phi$, the athermal cumulant ratio, introduced in \Sec{mic}, has a quite
simple form
\begin{equation}\EQ{propto} k(\alpha) \propto q \text{.} \end{equation}
An alternative parameterization, similar to \Eq{Efq}, can be obtained also for the
Langevin dynamics with exponential shot noise. For this, one needs merely to introduce
a location parameter into the compound Poisson process, considered in Appendix~\ref{sec:apx},
like we did in \Eq{muqf}. The constant of proportionality in \Eq{propto}, for
simple exponential noise, $\const_\mathrm{ex}$, and exponential shot noise, $\const_\mathrm{sh}$,
are
\begin{equation}
	\const_\mathrm{ex}= \frac{3}{16} (2 + \frac{b^2}{a^2});\quad
	\const_\mathrm{sh}= \frac{1}{4} (2 + \frac{b^2}{a^2}).
\end{equation}

\SEC{alg}{A macroscopically symplectic algorithm for simulations of Langevin Dynamics}
The differential equation~(\ref{eq:lgv2}) is mathematically equivalent to the following system:
\begin{equation}\EQ{lgv}
	\begin{cases}
		\dot\alpha(t) = \beta(t)\\
		\dot\beta(t) = - a \beta(t) - b^2 \alpha(t) + F + A\omega(t) + B\epsilon(t/\tau)\\
	\end{cases}.
\end{equation}
The macroscopic limit of \Eq{lgv} is
\begin{equation}\EQ{mac}
	\begin{cases}
		\dot\alpha(t) = \beta(t)\\
		\dot\beta(t) = - a \beta(t) - b^2 \alpha(t) + f\\
	\end{cases};
\end{equation}
cf. Eqs.~(\ref{eq:lgv1})-(\ref{eq:lgv2}) in \Sec{intro}. Equation~(\ref{eq:mac})
is time-reversible, like the equations of motion in our MD model of \Sec{mic}.

In order to design an algorithm for the LD simulations, which respects the macroscopic
time reversibility, consider first the deterministic equation~(\ref{eq:mac}). The phase space
of this system, as well as of \Eq{lgv}, is two-dimensional, $\pnt{\Gamma} = (\alpha\,\beta)$.
We are looking for a numerical scheme of the second order in time, which provides the accuracy level of our
MD simulations in \Sec{mic} and \Ref{PRE2016III}, by using the operator splitting formalism
\cite{TuckermanMartyna,MartynaTuckermanI,MartynaTuckermanII}; cf. \cite[Appendix C]{PRE2016III}.
Let us rewrite \Eq{lgv} as
\begin{equation}\EQ{dff} \dot{\pnt\Gamma}(t) = \Lvl \Gamma(t),\end{equation}
with the Liouvillian \cite[Chapter 3]{EvansMorriss}
$$ \Lvl = \beta\partial_\alpha - a\beta\partial_\beta - b^2 \alpha\partial_\beta + f\partial_\beta, $$
which is a sum of three mutually non-commuting terms
\begin{equation}\EQ{Lvls}
	\Lvl_\alpha = \beta\partial_\alpha;\quad
	\Lvl_{\beta\beta} = -a \beta\partial_\beta;\quad
	\Lvl_{\alpha\beta} = (f - b^2 \alpha)\partial_\beta.
\end{equation}
A formal solution of \Eq{dff} is
\begin{equation}\EQ{sln} \pnt{\Gamma}(t) = \exp(\Lvl t) \Gamma(0). \end{equation}

If we neglect for a moment the dissipative part of the dynamics in \Eq{sln}, $\Lvl_{\beta\beta}$,
we are left with two possible choices of a symplectic algorithm, described in \Ref{TuckermanMartyna}, namely,
a widely used velocity Verlet scheme,
\begin{eqnarray}\EQ{velV}
	\pnt\Gamma(t + \Delta{t}) &=& \exp(\Lvl_{\alpha\beta} \Delta{t}/2) \exp(\Lvl_\alpha \Delta{t})
		\exp(\Lvl_{\alpha} \Delta{t}/2)\nonumber\\
		&\times& \pnt\Gamma(t) + \mathcal{\pnt{O}}(\Delta{t}^3),
\end{eqnarray}
and a perhaps less known position Verlet scheme
\begin{eqnarray}\EQ{posV}
	\pnt\Gamma(t + \Delta{t}) &=& \exp(\Lvl_\alpha \Delta{t}/2) \exp(\Lvl_{\alpha\beta} \Delta{t})
		\exp(\Lvl_{\alpha} \Delta{t}/2)\nonumber\\
		&\times& \pnt\Gamma(t) + \mathcal{\pnt{O}}(\Delta{t}^3),
\end{eqnarray}
where $\Delta{t}$ is a time step of numerical integration.

Now we recall, that $\Lvl_{\alpha\beta}$ contains a force term $f$, which is the macroscopic
limit of the right-hand side in \Eq{lgv2}. Therefore, to obtain a mesoscopic version of Eqs.~(\ref{eq:velV})
and (\ref{eq:posV}), we make a substitution
\begin{equation}\EQ{macmes}
f \partial_\beta \Delta{t} \to \int_0^{\Delta{t}}dt[F + A \omega(t) + B \epsilon(t/\tau)] \partial_\beta
	= R(\Delta{t})\partial_\beta,
\end{equation}
where we used \Eq{Rt}.

In the Verlet algorithms above, \Eq{macmes} affects only an operator of the form $\exp(\Lvl_{\alpha\beta}\Delta{t})$;
cf \Eq{Lvls}. As explained shortly, its implementation involves sampling of random variables,
which is usually the most computationally expansive part of stochastic simulations. Note that,
on the right-hand side of \Eq{velV}, this operator is applied two times, in the form $\exp(\Lvl_{\alpha\beta}\Delta{t}/2)$,
while \Eq{posV} uses it only once. Therefore we continue with the latter integration scheme,
which is computationally more efficient.

By following the approach of Refs.~\cite{MartynaTuckermanI,MartynaTuckermanII}, the dissipative
part of the dynamics, due to $\Lvl_{\beta\beta}$, which was until now disregarded by \Eq{posV},
can be numerically implemented in two qualitatively equivalent ways. The one, which we
chose for the simulations in \Sec{mes}, is given by\footnote{
	If, everywhere in \Eq{sim}, we exchange the order, in which the operators $\exp(\Lvl_\alpha \Delta{t}/2)$ and $\exp(\Lvl_{\alpha\beta} \Delta{t}/2)$
	are applied, we would obtain the second implementation of the dissipative dynamics for
	the position Verlet scheme.
}
\begin{widetext}
\begin{eqnarray}\EQ{sim}
	\pnt\Gamma(t + \Delta{t}) =
		\exp\left(\frac{\Lvl_\alpha \Delta{t}}{2}\right) \exp\left(\frac{\Lvl_{\beta\beta} \Delta{t}}{2}\right)
		\exp(\Lvl_{\alpha\beta} \Delta{t})
	\exp\left(\frac{\Lvl_{\beta\beta} \Delta{t}}{2}\right) \exp\left(\frac{\Lvl_{\alpha} \Delta{t}}{2}\right)
	\pnt\Gamma(t) + \mathcal{\pnt{O}}(\Delta{t}^3).
\end{eqnarray}
\end{widetext}

The exact steps of the simulation algorithm can now be read from \Eq{sim}, as explained in
detail in Ref.~\cite[Appendix E]{FrenkelSmit}. On the right-hand side, we first operate on
$\pnt\Gamma(t)$ with $\exp(\Lvl_{\alpha} \Delta{t}/2)$, which generates a translation of $\alpha$
by $\beta \Delta{t}/2$ and leaves intact $\beta$. Then, from right to left, $\exp(\Lvl_{\beta\beta} \Delta{t}/2)$
scales $\beta$ by $\exp(-a \Delta{t}/2)$, while $\exp(\Lvl_{\alpha\beta} \Delta{t})$ increments it
by $b^2\alpha\Delta{t} + R(\Delta{t})$, and so on. In a compact form, a complete
step of the simulation, $\Delta{t}$, can be written as
\begin{widetext}
\begin{eqnarray}
\EQ{s1}
	\alpha(t+\Delta{t}/2) &=& \alpha(t) + \beta(t)\Delta{t}/2;\\
\EQ{s2}
	\beta(t+\Delta{t}) &=& \left[
		\exp\left(- \frac{a\Delta{t}}{2}\right)\beta(t) - b^2\alpha(t+\Delta{t}/2)\Delta{t} + R(\Delta{t})
	\right]\exp\left(- \frac{a\Delta{t}}{2}\right);\\
\EQ{s3}
	\alpha(t+\Delta{t}) &=& \alpha(t+\Delta{t}/2) + \beta(t+\Delta{t})\Delta{t}/2.
\end{eqnarray}
\end{widetext}

Note, that the above algorithm requires generation of the random variable $R(\Delta{t})$ only
in the second intermediate step, \Eq{s2}. When \Eq{lgv2} contains exponential noise, the athermal
part of $R(\Delta{t})$ is sampled once from a Gamma probability distribution \cite{PRE2016II}.
In the case of exponential shot noise, first, one has to generate a random integer from the
Poisson distribution. This number then determines how many exponentially distributed terms
must be sampled to evaluate $R(\Delta{t})$. For the whole duration of a LD simulation, on average,
exponential shot noise requires generation of $1 + \Delta{t}/\tau$ athermal random variates
per integration step $\Delta{t}$, while solely one is needed for simple exponential noise.

\bibliographystyle{apsrev4-1}
\bibliography{References}
\end{document}